\documentclass[aip,BMF,amsmath,amssymb,showpacs,floatfix,citeautoscript]{revtex4-1}

\usepackage{overpic}
 \usepackage[T1]{fontenc}
\usepackage{amsmath}
\usepackage{amssymb}
\usepackage{units}
\usepackage{xcolor}
\usepackage{graphicx}
\usepackage{upgreek}
\usepackage{mathrsfs}

\newcommand*{\ellp}{\ell_{\rm P}}
\newcommand*{\lsa}{\ell_{\rm SA}}
\newcommand*{\weff}{w} 

\newcommand{\dhh}{D_{\rm H}}
\newcommand*{\dw}{D_{\rm W}}

\usepackage{xcolor}

\DeclareMathOperator\erf{erf}

\begin{document}
\title{Hairpins in the conformations of a confined polymer}
\author{E. Werner}
\affiliation{Department of Physics, University of Gothenburg, Origov\"{a}gen 6B, 412 96 G\"{o}teborg, Sweden}
\author{A. Jain}
\affiliation{Department of Chemical Engineering
and Materials Science, University of Minnesota -- Twin Cities,
421 Washington Avenue SE, Minneapolis, Minnesota 55455, USA}
\author{A. Muralidhar}
\affiliation{Department of Chemical Engineering
and Materials Science, University of Minnesota -- Twin Cities,
421 Washington Avenue SE, Minneapolis, Minnesota 55455, USA}
\author{K. Frykholm}
\affiliation{Department of Biology and Biological Engineering, Chalmers University of Technology, 412 96 G\"{o}teborg, Sweden}
\author{T. St Clere Smithe}
\affiliation{Department of Physics, University of Gothenburg, Origov\"{a}gen 6B, 412 96 G\"{o}teborg, Sweden}
\author{J. Fritzsche}
\affiliation{Department of Physics, Chalmers University of Technology, 412 96 G\"{o}teborg, Sweden}
\author{F. Westerlund}
\affiliation{Department of Biology and Biological Engineering, Chalmers University of Technology, Kemiv\"agen 10, 412 96 G\"{o}teborg, Sweden}
\author{K. D. Dorfman}
\affiliation{Department of Chemical Engineering
and Materials Science, University of Minnesota -- Twin Cities,
421 Washington Avenue SE, Minneapolis, Minnesota 55455, USA}
\author{B. Mehlig}
\affiliation{Department of Physics, University of Gothenburg, Origov\"{a}gen 6B, 412 96 G\"{o}teborg, Sweden}

\begin{abstract} 
 If a semiflexible polymer confined to a narrow channel bends around by 180 degrees, the polymer is said to exhibit a hairpin. The equilibrium extension statistics of the confined polymer are well understood when hairpins are vanishingly rare  or when they are plentiful. Here we analyze the extension statistics in the intermediate situation via experiments with DNA coated by the protein RecA, which enhances the stiffness of the DNA molecule by approximately one order of magnitude.  We find that the extension distribution is highly non-Gaussian, in good agreement with Monte Carlo simulations of confined discrete wormlike chains.  We develop a simple model that qualitatively explains the form of the extension distribution. The model shows that the tail of the distribution at short extensions is determined by conformations with one hairpin.
\end{abstract}

\maketitle

\section{Introduction}
A semiflexible polymer confined to a very narrow channel is almost perfectly extended and aligned with the channel axis. Small deviations from perfect alignment yield an extension slightly smaller than the contour length of the polymer \cite{odijk1983}. In this so-called Odijk regime, the equilibrium statistics of the polymer extension approach a Gaussian distribution in the limit of large contour length. The mean and variance of this distribution are known to high precision \cite{burkhardt2010}. For a wormlike chain with persistence length $\ellp$ confined to a wider rectangular channel with largest side length $\dw$, this regime is obtained when $\dw \ll \ellp$ \cite{odijk2008,muralidhar2014a,werner2015}. 

\begin{figure}[b]
\mbox{}\hspace*{-8cm}
\begin{overpic}[width=8cm]{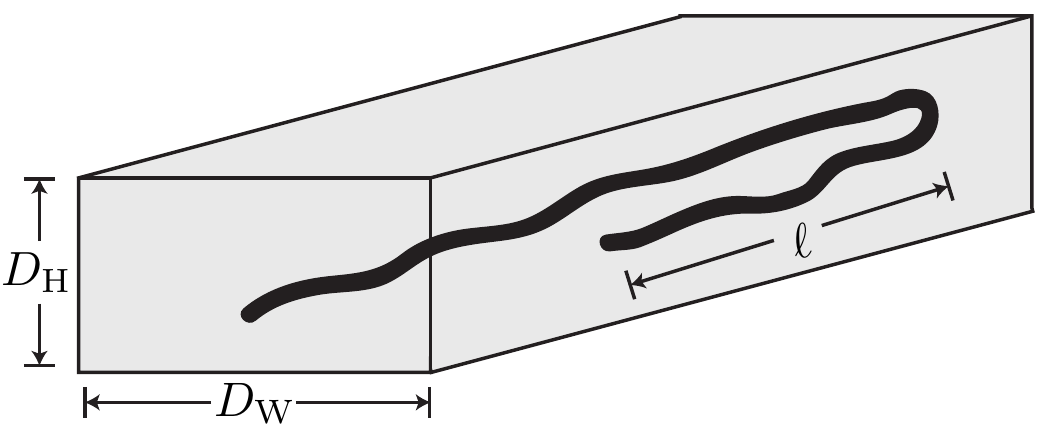}
\end{overpic}
\caption{\label{fig:hairpinSketch} Schematic of a confined wormlike chain confined in a channel of width $\dw$ and height $\dhh < \dw$, showing a configuration with a single hairpin of length $\ell$.}
\end{figure}

For wider channels ($\dw\approx \ellp$), the polymer can easily turn around, forming a C-shaped `hairpin' 
of length $\ell$ as illustrated in Fig.~\ref{fig:hairpinSketch}. As a result, the conformational statistics depend on the interaction between hairpin segments. 
Commonly, the strength of interaction between segments in self-avoiding polymers is  parameterized by an effective width, $w$.
The limit $w=0$ corresponds to an ideal polymer where self-avoidance does not matter. In this case the typical length of a hairpin defines a length scale $g$, the global persistence length \cite{odijk2006dna,odijk2008}. If the contour length $L$ is much larger than $g$ 
there are multiple hairpins in any channel segment, and the distribution of the extension approaches that of a one-dimensional random walk of $\approx L/g$ steps of length $\approx g$ \cite{odijk2008,muralidhar2014a}. 
What happens when $w$ is not zero?
The importance of the parameter $w$ is quantified by Odijk's scaling parameter
\begin{equation}
\label{eq:xi}
\xi = \frac{gw}{\dhh\dw^{2/3}\ellp^{1/3}},
\end{equation}
where the channel height $\dhh$ is assumed to be smaller than the channel width, $\dhh \le \dw$.
This parameter measures the expected number of overlapping points between the two strands of a hairpin of length $g$ \cite{odijk2008,muralidhar2014a,werner2015}.
The effect of self-avoidance depends on the magnitude of $\xi$.
If $\xi \gg 1$, hairpins are rarely observed, as it would be very difficult for the two strands of a hairpin to avoid overlapping with each other. As a result, in this regime the extension statistics are in approximate agreement with the predictions of the Odijk regime \cite{odijk2008,muralidhar2014a}.
If, on the other hand, $\xi \ll 1$, then self-avoidance has a negligible effect on the likelihood of forming a hairpin. Yet for a long polymer the effect on macroscopic observables such as the extension can be very significant. For example, while the average extension of a long ideal polymer ($w=0)$ grows as $\langle X \rangle \sim \sqrt{L}$, for a self-avoiding polymer the scaling of the extension is always linear in $L$ in the limit $L\to\infty$, 
regardless of how small $w>0$ is.

The results summarized above apply provided that $L\gg g$. 
However, since the global persistence length $g$ increases exponentially
as a function of $\ellp/D$ for large values of $\ellp/D$ 
\cite{odijk2006dna,muralidhar2014a,Chen2017}, the inequality $L \gg g$ 
can be violated even for very long polymers that are strongly confined or
have a large persistence length, so that $\ellp/D$ is large. 

In this paper we describe experiments with nanoconfined DNA that is 
covered with the protein RecA. The experimental setup is similar to the one in Ref.~\citenum{frykholm2014}. The persistence length of the DNA-RecA filament is an order of magnitude larger than that of bare DNA. We measured the extension distributions of DNA-RecA filaments in nanochannels of different widths,
ranging from $600\,$nm to $3\,\mu$m, and found that the equilibrium extension distribution is highly non-Gaussian. This conclusion is supported 
by direct numerical simulations of confined discrete wormlike chains. Comparison between experiments and simulations indicates
that the persistence length of the DNA-RecA filaments is approximately $\ellp = 2\,\upmu$m (this value is larger than most previous measurements reported in the literature \cite{frykholm2014,Hegner1999}).
The simulations also show that the global persistence length $g$  is larger than the contour length $L$ of the filaments.
This leads us to conclude that the extension of the filament in the wider channels is dominated by hairpin configurations 
such as that shown in Fig.~\ref{fig:hairpinSketch}. Because $g > L$, none of the theories summarized above apply in our case. 
We therefore formulate a simple model that qualitatively explains the shape
of the extension distributions.  The model relies on an expansion of the equilibrium distribution into a series of terms corresponding to conformations with no or one hairpin. We find that the main peak is determined by conformations without hairpins, and that the tail of the distribution at short extensions is due to conformations with one hairpin.

\section{Methods}
\subsection{Experiments}
\label{sec:experimentalDetails}
Our experiments used nanochannels with rectangular cross sections, fabricated as described in 
Ref.~\citenum{Persson2010}. 
The channel width $\dw$ ranged from $600\,$nm to $3000\,$nm, the channel height was $\dhh=140\,$nm. The channels were passivated using a lipid bilayer as previously described \cite{persson2012lipid}. Double-stranded DNA, either $\lambda$-DNA (New England Biolabs) or T4-DNA (Wako Chemicals), was coated with fluorescently labeled RecA protein as described in Ref.~\citenum{frykholm2014}. The resulting DNA-RecA filaments were moved between channels of different
widths using pressure-driven flow. In this way, we could observe {\em the same} DNA-RecA filament 
in channels of different widths $\dw$.
After moving the filament into a given channel, fluorescence imaging of the filament was performed while uniform pressure was maintained in the nanochannel. 
We used a Zeiss Axiovision microscope equipped with a $100\,$W mercury lamp, a Photometrics Evolve EMCCD camera, and a $100\times$ oil immersion TIRF objective (NA = $1.46$) from Zeiss. One video of $400$ frames was recorded for each channel width, for each filament. The time interval between frames was 0.11 seconds. The contour lengths of the T4-DNA-RecA filaments ranged from $7\,\upmu$m to $21\,\upmu$m, and those of the $\lambda$-DNA-RecA filaments ranged from $11\,\upmu$m to $23\,\upmu$m. 

\begin{figure}[t]
\mbox{}\hspace*{-8cm}
\begin{overpic}[width=8cm]{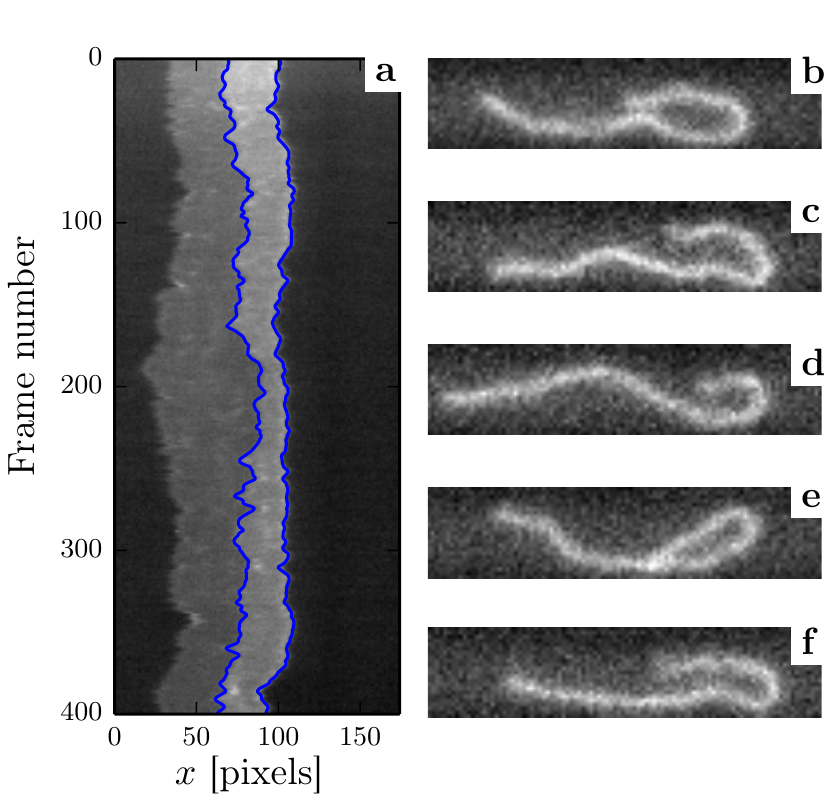}
\end{overpic}
\caption{\label{fig:kymographs}
{\bf a} Kymograph, $\lambda$-DNA coated with RecA, channel width $\dw = 2.4\upmu$m. Blue lines indicate the edges of the hairpin, as identified by the algorithm described in Section~\ref{sec:experimentalDetails}.
Panels {\bf b} to {\bf f} show cropped microscope images of the confined 
DNA-RecA filament (frame numbers $1$, $100$, $200$, $300$ and $350$). }
\end{figure}

We measured the span of the DNA-RecA filaments as an estimate of the extension (as opposed to the end-to-end distance).
To extract the span as a function of time, we fitted the brightness values for each frame to the curve $\alpha + \beta (\erf[\gamma(x - x_0)] - \erf[\varepsilon(x - x_1)])$, where $x$ is the location along the channel, and $\alpha,\beta,x_0$ and $x_1$ are fitting parameters \cite{tegenfeldt2004}. 
Kymographs (Fig.~\ref{fig:kymographs}) were produced by stacking the resulting intensity profiles into columns, 
so that each row of the column represents a single frame of the experimental recording.
To produce the intensity profiles, the section of the frame containing the molecule was identified by locating the region with the maximal brightness. Next, the pixel intensity was averaged over the direction perpendicular to the channel, resulting in a row of pixel values which we interpret as the intensity profile along the channel. 

In practice, we found that we could obtain somewhat more robust results for the extension by modifying the algorithm for computing the kymographs. Instead of creating the intensity profile by averaging across the channel, we first smoothed each video frame by a median filter of radius two pixels, and defined the profile as the maximum pixel intensity in each column. Further, before fitting to the box curve described above, we smoothed the resulting kymograph by a moving time average with a window size of three frames and a moving spatial
average with a window size of four pixels. Kymographic representations of all experimental videos are available as supplementary material.

\begin{figure*}
\begin{overpic}[width=\textwidth]{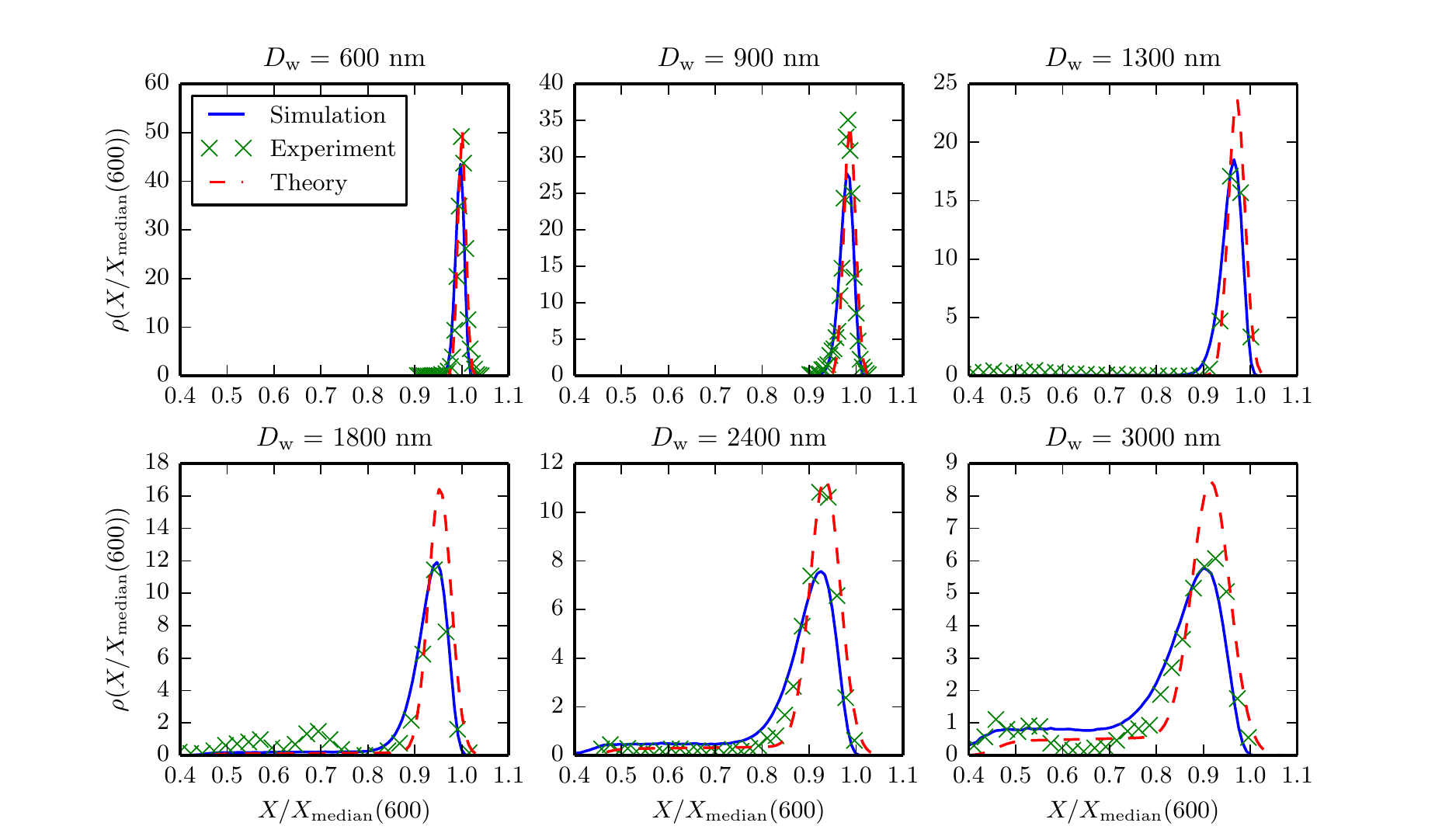}
\end{overpic}
\caption{\label{fig:extension_simulationAndExperiment} Distributions of $\rho(X/X_{\rm median}(600))$ of the span $X$ relative to the median span $X_{\rm median}(600)$ in the 600 nm channel from computer simulations (solid lines) and experiments (symbols). Simulation
parameters $L=15\,\mu$m and $\ellp=2 \,\mu$m.
Since we cannot measure the contour length to sufficient accuracy in experiments, we instead normalize the span by dividing by the median extension in the narrowest channel ($\dw=600\,$nm). The simulations yield $X_{\rm median}(600) \approx 14.18\,\upmu$m. The experimental data consist of the subset of experiments for which the same filament was measured both in the channel under consideration and in the narrowest channel ($\dw=600\,$nm), and further where $X_{\rm median}(600)$ agrees to within 20\% with the value found in simulation. Further, we removed two filaments that exhibited a hairpin in the narrowest channel, in which case the median span is not a good proxy for the contour length.  Also shown are the model predictions  (Section \ref{sec:theory}), dashed lines, for $L=15\,\upmu$m and $\ellp=2 \,\upmu$m. For the  narrowest channel we assume $g=\infty$ in the theory, i.e. no hairpins. 
}
\end{figure*}

\subsection{Direct numerical simulations}
Our equilibrium simulations used
a discretized wormlike chain model \cite{wang2005}.
The discretization was achieved using a touching-bead
model, with $N+1$ beads of diameter $w$ connected by rigid bonds. The
bending potential 
\begin{equation}
\label{eq:U}
U(\theta_1, \dots, \theta_N) = k_B T \kappa \sum_{n=1}^{N} 
(1 - \cos \theta_n)\,.
\end{equation}
was used to model the stiffness of the chain.
In Eq.~(\ref{eq:U}), the index $n$ ranges over bonds, 
$\theta_n$ is the angle between bonds $n$
and $n+1$, and $\kappa$ is the bending constant. The persistence length in this model is defined as \cite{lodgeBook,Dorfman2017}
\begin{equation}
\ellp = w \sum_{k=0}^{\infty} \langle \mathbf{t}_n  \cdot \mathbf{t}_{n+k} \rangle\,,
\end{equation}
where $\mathbf{t}_n$ is the tangent vector to the chain at bond
$n$, and the average is computed for an ideal and unconfined chain. 
The persistence length can be expressed in terms of $\kappa$
and $w$ as  \cite{muralidhar2015}
\begin{equation}
\label{eq:lp}
\frac{\ellp}{w} = \frac{\kappa }{\kappa - \kappa \coth \kappa+1}
\approx \kappa
\end{equation}
for large values of $\kappa$.
Details of the definition
of $\ellp$ and the derivation of Eq.~(\ref{eq:lp}) are given
in Ref.~\citenum{Dorfman2017}.

\begin{table}[b]
\mbox{}\hspace*{-6.5cm}
\begin{tabular}{ccccccccc}
\hline\hline
$L$ & $\ellp$ & $w$  &$\dhh$&&  $\dw$ \\\hline
                15 $\upmu$m & 2  $\upmu$m & 10 nm & 140 nm && 0.6, 0.9, 1.3, 1.8, 2.4, 3.0  $\upmu$m \\\hline\hline\\
\end{tabular}
\caption{\label{tab:1} Parameters used in the computer 
simulations of the DNA-RecA filaments. The contour length used in the simulations is a representative value for the DNA-RecA filaments observed in experiments such as those in Fig.~\ref{fig:hairpinSketch}. The value of the persistence length was selected by comparison between simulation data and experiment {(see text)}. The value of the effective width represents an approximation for the thickening of naked DNA from coating with RecA.}
\end{table}
Excluded-volume interactions were incorporated into the model by imposing an infinite energy barrier for bead-bead overlap and bead-wall overlap. 
We simulated the model using the pruned-enriched Rosenbluth method (PERM) \cite{grassberger1997,Prellberg2004}, following the approach described in our previous work\cite{tree2013,Tree:2013a}. PERM is a biased chain growth method that provides, {\it inter alia}, information on the extension statistics while avoiding the attrition problem in self-avoiding random walks. The PERM simulations were conducted to grow chains up to 1501 beads using the parameters given in Table \ref{tab:1}. Distributions of the mean span were obtained from $2 \times 10^6$ tours, corresponding to $1.7 \times 10^7$ configurations.

\section{Results and discussion}
\subsection{Extension distribution - experiments and simulations }
\label{sec:compareToExp_equilibrium}
Figure~\ref{fig:extension_simulationAndExperiment} compares the distribution of the extension $X$ of the 
DNA-RecA filament along the channel direction measured in experiments against the results from simulations.
It is customary to normalize the extension $X$ by dividing by the contour length $L$ of the polymer.
However, since  it is hard to determine the experimental contour length precisely enough, we normalized the polymer extension by the median polymer extension in the narrowest channel 
$\dw=600\,$nm.
We must therefore restrict our analysis to the subset of filaments that were imaged in the narrowest channel. 

\begin{figure}
\mbox{}\hspace*{-2cm}
\begin{overpic}[width=14cm]{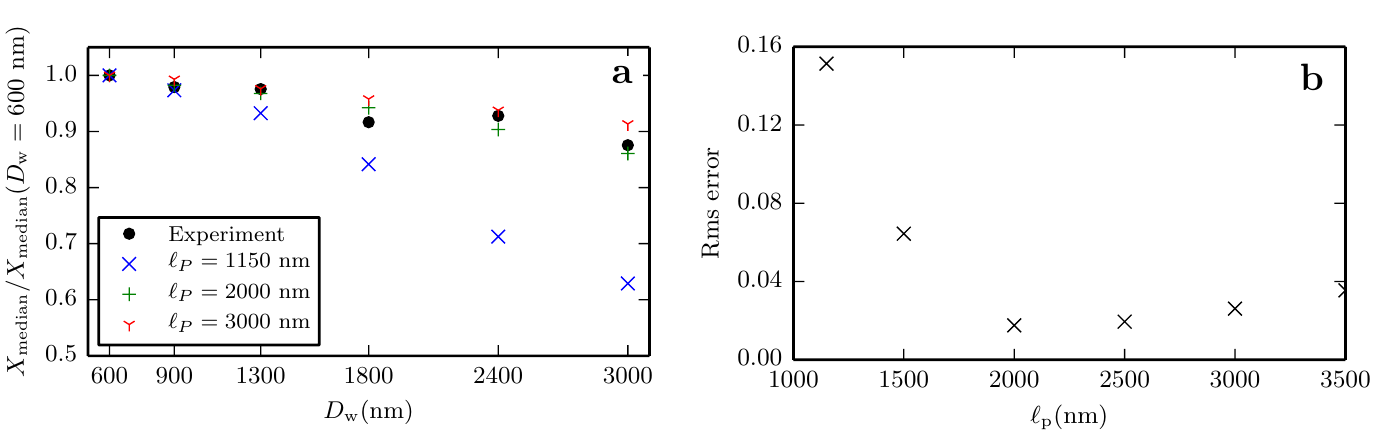}
\end{overpic}
\caption{\label{fig:determine_lP}
Comparison of simulations against experiments, to establish which value of the persistence length $\ellp$ fits best. {\bf a} Comparison of the median 
DNA-RecA filament span from experiments against simulations for different values of $\ellp$. The median span is normalized by the value measured in the narrowest channel (as in Fig.~\ref{fig:extension_simulationAndExperiment}).
{\bf b} Root-mean squared difference between the simulated and experimental values shown in panel {\bf a}, averaged over all channels except the smallest ones (where the difference is zero by construction).
}
\end{figure}

To compare simulations against experiments,  it is necessary to determine the value of the persistence length $\ellp$ that best describes the DNA-RecA filament. To this end, we measured the median extension for each channel size, and compared the results against simulations for different values of $\ellp$. The results are shown in Fig.~\ref{fig:determine_lP}. We observe the best fit between simulations and experiment when $\ellp$ is close to 2$\,\upmu$m. We therefore used this value  in Fig.~\ref{fig:extension_simulationAndExperiment}.
Fig.~\ref{fig:determine_lP} shows  that the value  $\ellp=1.15\,\upmu$m quoted  in Ref.~\citenum{frykholm2014} is not consistent with our simulation data at large values of $\dw$. In Ref.~\citenum{frykholm2014}, the persistence length was determined  by fitting Odijk's expression Eq.~(\ref{eq:Odijk_mean}) to  the experimental data, using $L$ and $\ellp$ as fitting parameters. This procedure is quite insensitive to $\ellp$ and does not allow to rule out $\ellp=2\,\upmu$m (Appendix \ref{app:C}).

Our estimate of the persistence length  comes with a large uncertainty,  because the minimum in Fig.~\ref{fig:determine_lP}(b) is quite shallow. However, our estimate appears to be inconsistent with the  value $\ellp\approx 1\,\upmu$m found in Ref.~\citenum{Hegner1999} obtained using an entirely different technique. 
At present we do not know the reason for this discrepancy.
 One previous study \cite{loenhout_dynamics_2009}, on the other hand, found a persistence length of $2.1 \pm 0.1\,\upmu$m for single-stranded DNA covered with RecA, albeit under experimental conditions that differ from ours. 

\begin{table}
\mbox{}\hspace*{-10cm}
\begin{tabular}{lcccccc}
&&&&&&\\\hline\hline
$\dw$ ($\upmu$m) & 0.6 & 0.9 & 1.3 & 1.8 & 2.4 & 3.0\\
\hline
$g$ ($\upmu$m) & --- & 13000 & 350 & 90 & 37 & 22\\
$\xi$ & --- & 780 & 17 & 3.5 & 1.2 & 0.61 \\
$\lsa$ ($\upmu$m) & ---   &  9.4      & 12 & 15 & 18 & 21 \\
\hline
\hline
\end{tabular}
\caption{\label{tab:2} Summary of simulation data and corresponding theory parameters for $L=15\,\mu$m. The values of $g$ are obtained from simulations of ideal chains for the parameters in Table \ref{tab:1}, as described in Ref.~\citenum{muralidhar_backfolded_2016}.  The entries ---  indicate channel sizes where $g$ could not be measured. The values of the scaling parameter $\xi$ are computed from Eq.~(\ref{eq:xi}). The values of the self avoidance length scale $\lsa$ are obtained from Eq.~(\ref{eq:P_noOverlap}) using the value of $\alpha$ computed from Fig.~\ref{fig:hairpinLengthDistribution}. }
\end{table}

Figure~\ref{fig:extension_simulationAndExperiment} shows that there is good qualitative agreement between the simulations for  $\ellp=2\,\upmu$m and the experiments.
This indicates that the discretized wormlike chain model is a good approximation for the conformations of nanoconfined DNA-RecA filaments. In addition
to the main peak, i.e., the peak located near $X/X_{\rm median}(600) \approx 0.9$, the distributions in the wider channels exhibit pronounced tails at short spans. Using the computer simulations we verified that the small-$X$ tails correspond to conformations
with at least one hairpin (as depicted in Fig.~\ref{fig:hairpinSketch}).

Figure~\ref{fig:extension_simulationAndExperiment} also shows quantitative differences between simulations and experiments. These differences may be due to the fact that the DNA-RecA filament conformations may have been insufficiently sampled in the experiments.  The experimentally observed conformation dynamics (discussed below) indicate that the main peak in the extension distribution is likely to be well sampled, but the tails at short filament extensions (spans) may not be well sampled.
Another possible source of error is that the experimental method may have caused a bias in the initial conformations, because the filaments were moved between different regions of the channels between the measurements, from one channel size to a narrower one.  However, we expect this bias to be relatively small in our case when compared to injection from a reservoir into a nanoslit\cite{Muthkumar}, since the change in confinement between two values of $\dw$ is relatively small.  We have insufficient data to evaluate a possible bias with statistical analysis, but we checked at least that in none of the cases analyzed in this Section  did a filament ending in a left (right) hairpin state in one channel begin with a left (right) hairpin in the next channel. This indicates that the initial conditions were in equilibrium.

Since only the protein RecA fluoresces, parts of DNA that are not coated by RecA are invisible in our experiments. 
It is therefore possible that such invisible DNA is attached at one or both ends of the visible filament. This could have a small effect on the equilibrium statistics, as excluded volume interactions between the bare and coated DNA could reduce the probability of observing hairpins.  More importantly, hydrodynamic friction between the bare DNA and the surrounding solution might slow down the dynamics.  

\subsection{Global persistence length}
\label{sec:simulations}
According to Eq.~(\ref{eq:xi}), it is the parameter $\xi$ (the scaled global persistence length)  that determines how frequently hairpins can form. Therefore
we computed this parameter in our simulations.
Odijk \cite{odijk2006dna,odijk2008} proposed a mechanical theory for computing the global persistence length $g$. However, recent simulations of confined wormlike chains suggest that Odijk's theory overestimates the global persistence length \cite{muralidhar2014a,muralidhar2016,muralidhar_backfolded_2016}. While the functional form predicted by Odijk's theory is reasonable for square and circular channels \cite{muralidhar2014a,muralidhar2016}, it is not for rectangular channels \cite{muralidhar_backfolded_2016}. Therefore we decided to determine the values of $g$ for our particular channels via simulation,  following the approach for rectangular channels
described in Ref.~\citenum{muralidhar_backfolded_2016}. 

Table~\ref{tab:2} presents the values of $g$ obtained in this way. For channels with $\dw = 900$ nm and wider, we could reliably extract a value of the global persistence length. For the smallest channel, we could not compute a value of $g$ from the simulations. This suggests that $g$ is larger than the maximal contour length that we could simulate. In principle, this limitation can be surmounted by increasing the contour length of the simulated ideal chains. In practice, we expect that $g$ increases exponentially with decreasing $\dw/\ellp$ \cite{odijk2006dna}. As a result, the requisite contour length quickly becomes infeasible to simulate, even using a biased-growth method like PERM. In what follows, we simply set $g = \infty$ for the smaller channels. 
At any rate, Table \ref{tab:2} shows that $g\gg L$ for all but the widest channel.

Table~\ref{tab:2} also gives the corresponding values of the parameter $\xi$ in Eq.~(\ref{eq:xi}). We see that the values of $\xi$ are of order unity for the wider channels.
This is consistent with the shapes of the extension distributions that  exhibit pronounced tails at short spans due to hairpin-conformations for these channels. In the following we describe a simple model that qualitatively explains the shapes of the distributions in this limit.

\subsection{Model calculation}
\label{sec:theory}
When the parameter $\xi$ is of order unity,  the conformations may sometimes exhibit hairpins, and sometimes not. Since the mechanisms determining the polymer span $X$ are quite different if hairpins are present or absent, it is necessary to consider these possible configurations separately, before combining them into a single distribution $\rho(X)$:
\begin{equation}
\label{eq:distributionAsSumOverHairpins}
\rho(X) = \sum_{j=0}^\infty \rho_j(X) P_j\,.
\end{equation}
Here $\rho_j(X) \equiv \rho(X|N_{\rm h}=j)$ is the distribution of $X$
conditional on that the number $N_{\rm h}$ of hairpins equals $j$, and $P_j\equiv {\rm prob}(N_{\rm h}=j)$ is the probability of observing $j$ hairpins. Since this study is concerned with the situation where hairpins are rare, we restrict the following discussion to the two cases of either zero or one hairpin.

When $L\gg \lambda \approx \dw^{2/3} \ellp^{1/3}$, the distribution $\rho_0(X)$ for a polymer without hairpins is Gaussian
\begin{align}
\rho_0(x) &= (2\pi\sigma_{\rm Od}^2)^{-1/2} \exp\Big[-\frac{(X-\mu_{\rm Od})^2}{2\sigma_{\rm Od}^2}\Big]
\label{eq:rho0}
\end{align}
with mean and variance which we  approximate by the conformation statistics of the Odijk regime \cite{burkhardt2010},
\begin{align}
\label{eq:Odijk_mean}
\mu_{\rm Od} &= L\Big(1 - 0.091 \frac{\dhh^{2/3} + \dw^{2/3}}{\ellp^{2/3}}\Big), \\
\sigma^2_{\rm Od} &= 0.0048 L\Big(\frac{\dhh^2 + \dw^2}{\ellp}\Big).
\label{eq:Odijk_variance}
\end{align}
In Appendix \ref{app:A} we derive an expression for $\rho_1(X)$:
\begin{align}
\label{eq:rho1SA}
\rho_1(X) &{=\mathscr{N}} \Big[\erf\Big(\frac{2X - \mu_{\rm Od}}{2\sigma_{\rm Od}}\Big) - \erf\!\Big(\frac{X - \mu_{\rm Od}}{\sqrt{2}\sigma_{\rm Od}}\Big) \Big] \exp\Big(\frac{X-L/2}{\lsa}\Big)\,.
\end{align}
The calculation in Appendix \ref{app:A} assumes that hairpins are equally likely to form anywhere on the 
ideal polymer, that each of the ideal hairpin strands fluctuates according to Eqs.~(\ref{eq:rho0}-\ref{eq:Odijk_variance}). Self avoidance is taken into account by penalizing hairpins with long hairpin lengths $\ell$. We assume 
that the probability of collisions between hairpin strands is proportional to the length of the shorter strand.
The prefactor $\mathscr{N}$ in Eq.~(\ref{eq:rho1SA}) is determined
by the condition that $\rho_1(X)$ is normalized to unity on $[0,L]$.
The length scale $\lsa$  quantifies the hairpin length at
which self-avoidance becomes important, penalizing configurations with large $\ell$. The probability of overlaps between  two hairpin strands can be estimated by a mean-field argument \cite{odijk2008,muralidhar_backfolded_2016}, yielding
\begin{equation}
\label{eq:P_noOverlap}
P_{\rm no\; overlap}(\ell) = \exp\left(-\alpha\xi\ell/g\right)\equiv\exp(-\ell/\lsa)\,,
\end{equation}
where $\alpha$ is a prefactor of order unity.
 We estimated the parameter $\lsa$ from our computer simulations (Appendix \ref{app:B}). The results
are shown in Table~\ref{tab:2}. This parameter determines the ratio of the coefficients $P_1$ and $P_2$
in Eq.~(\ref{eq:distributionAsSumOverHairpins}). We show in Appendix \ref{app:A} that
\begin{equation}
\label{eq:P1P0}
\frac{P_1}{P_0} =
\frac{ \lsa}{g}\Big[1-\exp\Big(-\frac{L}{2\lsa}\Big)\Big] \,.
\end{equation}
As expected, the limit $\lsa \gg L$ reproduces the ideal result $({P_1}/{P_0})_{\rm ideal} = L/(2g)$ (Appendix \ref{app:A}). The opposite limit yields a ratio that is independent of $L$. 

In summary, Eqs.~(\ref{eq:distributionAsSumOverHairpins}), (\ref{eq:rho0}), (\ref{eq:rho1SA}), and (\ref{eq:P1P0}) yield the desired approximation for the distribution
$\rho(X)$ of the span $X$. 
The result is shown in Fig.~\ref{fig:extension_simulationAndExperiment} as dashed lines.
We observe good qualitative agreement with both experiments and simulations, without any fitting parameters.  For wider channels, the theory underestimates the width of the peak at large span, as well as its skewness. This is a result of the assumption that Odijk theory [Eqs.~(\ref{eq:Odijk_mean})--(\ref{eq:Odijk_variance})] describes the statistics of each hairpin, but this assumption starts to fail when $\dw\approx \ellp$.
We also note that the simple theory neglects the effect of the curved part of the hairpin on the extension (span)
of the filament in the channel. This results
in an error of the order of the channel width, so that the theory may overestimate the span by about $20\%$. 
Comparing simulation and theory in Fig.~\ref{fig:extension_simulationAndExperiment}, we see that the left tail does extend to smaller values of $X$ and is heavier in the simulations, compared with the theory. 

\subsection{Diffusion model for hairpin size}
\label{subsec:hairpinDynamics}
\begin{figure}[t]
\mbox{}\hspace*{-9cm}\mbox{}\begin{overpic}[width=7cm]{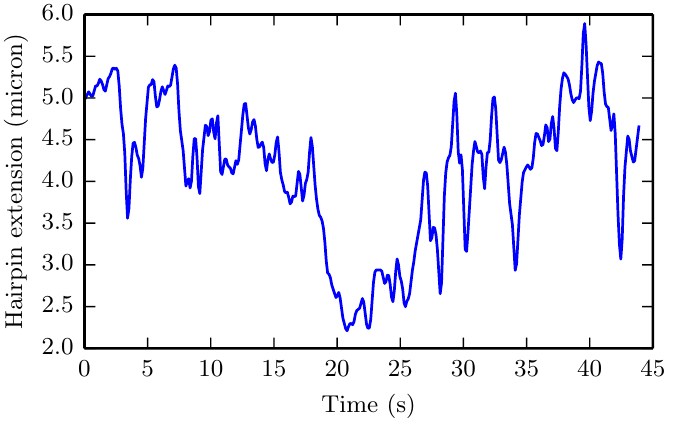}
\end{overpic}
\caption{\label{fig:hairpinDiffusion}
Span of the hairpin shown in Fig.~\ref{fig:kymographs} as a function of time. Note that the span
decreases significantly before increasing again, showing that the hairpin extension dynamics are
non-deterministic. }
\end{figure}
We mentioned above that a possible explanation for the difference between experiments and simulations in Fig.~\ref{fig:extension_simulationAndExperiment} is that the experimental
conformations have been insufficiently sampled. To investigate whether this is the case, we analyzed the experimental time series of the filament spans. A representative
result is shown in Fig.~\ref{fig:hairpinDiffusion}.
It demonstrates that the hairpin length first shrinks and then grows, demonstrating that the dynamics are stochastic, unlike the unfolding process described in 
Refs.~\citenum{levy_entropic_2008} and \citenum{ibanez-garcia_hairpin_2013}. 
To qualitatively explain this observation, and to estimate the unfolding time scale, we formulated a diffusion model for the hairpin dynamics.

A hairpin configuration can be represented by two polymer strands, connected at one end. If one of the strands is much longer than the other it is practically immobile. 
The shorter strand diffuses like a polymer in the Odijk regime, with diffusion constant $k_{\rm B}T/\zeta(\ell)$ \cite{muralidhar2015}. Here $\zeta$ is a friction coefficient. It is defined in the following way: Imagine that the polymer is dragged slowly along the channel with a velocity $v$. The opposing force acting on the polymer is then proportional to $v$, with proportionality constant $\zeta$. Since a change in the span requires that a large fraction of the polymer segments move, we can assume that $v \approx \dot{X}$. It follows that a change in the span leads to an average friction force $F \approx - \zeta \dot{X}$. The length of the hairpin changes as it diffuses. In addition to the diffusive motion, the strand experiences an unfolding force $f$, due to the collisions between the two strands. 
The simultaneous action of diffusion and the constant unfolding force can be described by the generalized diffusion equation
\begin{align}
\label{eq:fokkerPlanck}
\frac{\partial \rho(\ell)}{\partial t}\! = \!-\frac{\partial }{\partial \ell} \left[\frac{f}{2\zeta(\ell)}\rho(\ell)\right] \!+\! k_{{\rm B}} T\frac{\partial}{\partial \ell} \left[\frac{1}{2\zeta(\ell)} \frac{\partial\rho(\ell)}{\partial \ell} \right].
\end{align}
Note that Eq.~(\ref{eq:fokkerPlanck}) has the equilibrium distribution $\rho_{\rm eq}(\ell) \propto \exp[\int^\ell\! f/(k_{\rm B}T){\rm d}\ell']$, in agreement with the Boltzmann distribution if the force is given by the derivative of an energy, $f = -\partial E/\partial \ell $. In the deterministic limit, where the rightmost term is negligible, Eq.~(\ref{eq:fokkerPlanck}) is consistent with the deterministic description of an unfolding process given in Refs.~\citenum{levy_entropic_2008} and \citenum{ibanez-garcia_hairpin_2013}, as well as with the description of ejection of a polymer out of a nanochannel \cite{milchev_ejection_2010}.

The force $f$ in Eq.~(\ref{eq:fokkerPlanck}) is due to the self-avoiding interaction between the two strands in the folded configuration that tends to unfold the hairpin. This can be modeled by an entropic force given as the slope of the free energy \cite{werner2014}. Apart from terms that are independent of the macroscopic configuration, the free energy is given by the sum of the elastic energy stored in the hairpin bend and an entropic free-energy contribution, due to
collisions between the strands, $-k_{\rm B}T \log P_{\rm no\; overlap}$. The first term does not the depend on the hairpin length, so Eq.~(\ref{eq:P_noOverlap}) shows 
that the entropic force is given by:
\begin{align}
\label{eq:entropicForce}
f &= k_{\rm B}T \frac{\partial \log P_{\rm no\; overlap}}{\partial \ell} = \frac{-k_{\rm B}T}{\lsa} 
= -\alpha k_{\rm B}T\frac{\weff}{\dhh \ellp^{1/3} \dw^{2/3}}.
\end{align}	 
While it is possible to provide a more accurate estimate of the entropic force, for example by using the methods of Ref.\citenum{Polson2017}, we cannot directly apply the results of the latter reference to our analysis. The confinement free energy of a high aspect ratio rectangular channel, which arises in part due to excluded volume, is different than in a circular tube
\cite{odijk2008,Muralidhar2016a}.

Eq.~(\ref{eq:fokkerPlanck}) allows one to define two time scales for a hairpin of length $\ell$, one diffusive time scale $\tau_{\rm diff} = \zeta(\ell) \ell^2/(k T)$ and one deterministic time scale $\tau_{\rm drift} = \zeta(\ell) \ell/f$. In drift-diffusion problems, the relative effects of drift to diffusion is captured by the ratio of these time-scales, namely the P\'eclet number $\mathrm{Pe} = f \ell/kT$. For a hairpin of length 3 $\upmu$m in a 1300 nm channel, we find $\mathrm{Pe} \approx 0.2$, suggesting that drift is a small effect. 
The prediction that the unfolding dynamics of a DNA-RecA hairpin is diffusive contrasts with the predominantly deterministic dynamics seen in experiments on bare DNA \cite{levy_entropic_2008,alizadehheidari2015c}. This prediction is validated by the experimental data shown in Fig.~\ref{fig:hairpinDiffusion}. 
Since only a small fraction of our experiments show hairpins we do not have enough data for a quantitative test of Eq.~(\ref{eq:fokkerPlanck}), or to estimate  $\tau_{\rm diff}$, but the simulation results of Ref.~\citenum{muralidhar2015} yield an estimate $\tau_{\rm diff} \approx 10$ s for a hairpin of length 3 $\upmu$m. 
This is of the same order as the length of an experimental video ($\approx 40$ s). As discussed in Section~\ref{sec:compareToExp_equilibrium}, this could indicate that the small-$X$ tails of the experimental distributions in Fig.~\ref{fig:extension_simulationAndExperiment} are insufficiently sampled.

\section{Conclusions}
In this article we investigated the  extension distribution of nanoconfined RecA-coated DNA. We compared the experimental distribution of the filament span with results of equilibrium Monte Carlo simulations of
strongly confined wormlike chains that are shorter than the global persistence length. In the wider channels, the experimental distributions are strongly non-Gaussian, in good qualitative agreement with our results from the Monte Carlo simulations.  We found the best fit between simulations and experiments if we assumed  that the DNA-RecA filaments have a persistence length of $\ellp\approx 2\,\upmu$m.
This value is larger than  $\ellp\approx 1\,\upmu$m found in Ref.~\citenum{Hegner1999}, obtained using an entirely different technique. The reason for this discrepancy is not known. 

Apart from the central peak, the most distinct feature of the non-Gaussian distributions is a heavy tail at short spans. We derived a simple model for the extension distribution that explains that
this tail is due to conformations that contain at least one hairpin. This also furnishes a possible explanation for the quantitative differences that we observed between the experimental and simulation data. It could be partly due to the fact that the small-$X$ tails of the experimental distributions are insufficiently sampled, as mentioned above.
In addition to the time scale for unfolding discussed in Section~\ref{subsec:hairpinDynamics}, our ability to sample the tail is also affected by  the time scale for folding, since folding involves a high energy penalty.
We expect that the sampling of the tails of the distribution is the largest source of discrepancy between experiment and simulations in our system.  At any rate, our results show that the equilibrium theory works quite well even though the actual experiment may not sample sufficiently.  A recent analysis of unfolding of DNA in a nanochannel \cite{Kevin} allows a similar conclusion, although for quite different parameters. Despite the fact that the dynamics is deterministic on the time scales of that experiment, Odijk's equilibrium theory works very well.
We formulated a stochastic model for the stochastic dynamics of a hairpin extension to estimate the time scale associated with this dynamics, indicating the the time scale for hairpin length fluctuations is indeed similar to the length of an experimental video.

More generally, the hairpin dynamics of the DNA-RecA filaments are very different than the corresponding dynamics of bare DNA. While the latter are almost deterministic \cite{levy_entropic_2008,alizadehheidari2015c}, the former are not. We therefore proposed a stochastic model for how the size of a hairpin fluctuates over time.
This model includes both the effect of self-avoidance (which tends to reduce the length of the hairpin), and the diffusion caused by thermal motion. Since self-avoidance is relatively weak for our system, the dynamics is dominated by the diffusive term. Our experimental data are in qualitative agreement with the stochastic model, but further experiments and modeling work are necessary to establish a quantitative description of the hairpin dynamics.

\vfill\eject

\section*{Supplementary Material}
See supplementary material for a summary of the experimental data in the form of kymographs.

\begin{acknowledgments}
The labeled RecA protein was a kind gift from Edwige B. Garcin and Mauro Modesti.
EW and BM wish to thank Stefano Bo and Tobias Ambj\"ornsson for discussions concerning Eq.~(\ref{eq:fokkerPlanck}).
Financial support from Vetenskapsr\aa{}det [Grants No. 2013-3992 (BM) and 2011-4324 (FW)] and from the National Institutes of Health [R01-HG006851(KDD)] is gratefully
acknowledged. The computational work was carried out in part using
computing resources at the University of Minnesota Supercomputing
Institute.
\end{acknowledgments}

\appendix
\renewcommand\thefigure{S-\arabic{figure}}
\setcounter{figure}{0}

\section{Derivation of $\rho_1(X)$}
\label{app:A}
The approximation (\ref{eq:rho1SA}) for $\rho_1(X)$ is derived in several steps. 
First, let $\ell$ denote the length of the shortest hairpin strand, as illustrated in Fig.~\ref{fig:hairpinSketch}. In this configuration, the span of the filament in the channel is, to a good approximation, given by the extension of the longest hairpin strand, which obeys Odijk statistics with a contour length of $L - \ell$. For ideal polymers, i.e. without interactions between the two strands of the hairpin, the hairpin length is drawn uniformly between $0$ and $L/2$. Thus
\begin{align}
\label{eq:rho1_ideal}
\rho_1^{\rm id}(X) &\approx \frac{2}{L} \int_0^{L/2} d\ell \rho_0(X;L - \ell) \approx \frac{1}{\mu_{\rm Od}}\! \left[\erf\!\left(\frac{2X - \mu_{\rm Od}}{2\sigma_{\rm Od}}\right) - \erf\!\left(\frac{X - \mu_{\rm Od}}{\sqrt{2}\sigma_{\rm Od}}\right) \right]\,. 
\end{align}
Here $\rho_0(X;L)$ is the  distribution of the span of a polymer segment of length $L$ that is free of hairpins [Eq.~(\ref{eq:rho0})].

Second, we must  modify Eq.~(\ref{eq:rho1_ideal}) to account for excluded volume. 
If the polymer is self-avoiding, then configurations with large $\ell$ are penalized. The probability of overlaps between these two strands can be estimated by a  mean-field argument \cite{odijk2008,muralidhar_backfolded_2016}, yielding
\begin{equation}
\label{eq:lsa}
P_{\rm no\; overlap}(\ell) = \exp\left(-\alpha\xi\ell/g\right)\equiv\exp(-\ell/\lsa)\,,
\end{equation}
where $\alpha$ is a prefactor of order unity. This is Eq.~(\ref{eq:P_noOverlap}) in the main text. The second
equality defines the self-avoidance length $\lsa$. 
 Eq.~(\ref{eq:lsa}) allows us to modify Eq.~(\ref{eq:rho1_ideal}) to take into account self-avoidance: the  integrand in Eq.~(\ref{eq:rho1_ideal}) must include an extra factor proportional to $[1- P_{\rm no\; overlap}(\ell)]$. Performing the integration then yields
\begin{align}
\label{eq:rho1_SA_E}
&\rho_1(X) {=\mathscr{N}} \Big[\erf\Big(\frac{2X - \mu_{\rm Od}}{2\sigma_{\rm Od}}\Big) - \erf\!\Big(\frac{X - \mu_{\rm Od}}{\sqrt{2}\sigma_{\rm Od}}\Big) \Big] \exp\Big(\frac{X-L/2}{\lsa}\Big)\,.
\end{align}
This is Eq.~(\ref{eq:rho1SA}) in the main text.  The prefactor is given by the condition that $\rho_1(X)$ is normalized on $[0,L]$.

Third, to determine the first two terms
in Eq.~(\ref{eq:distributionAsSumOverHairpins}) in the main text  we also require $P_0$ and $P_1$.
For ideal polymers, the formation of a hairpin bend at a given location is independent of how many other hairpins that have already formed. It follows that the number of hairpins is Poisson distributed, with a rate constant $\mu \propto L/g$. A precise calculation yields $\mu = L/(2 g)$. 
It follows that $P_1^{\rm id}/P_0^{\rm id} = L/(2 g)$. We also note that
$P_2^{\rm id}/P_0^{\rm id} = L^2/(8 g^2)$, so the contribution of two hairpins to $\rho(X)$
is small when $L\ll g$.  Now consider a self-avoiding polymer. It is less likely than an ideal polymer to exhibit hairpins, as only a fraction of all ideal configurations with a hairpin are free of overlaps. 
We denote this fraction by
\begin{align}
\langle P_{\rm no\; overlap}\rangle &\equiv \frac{2}{L} \int_0^{L/2} \!\!{{\rm d}\ell}\,P_{\rm no\; overlap}(\ell)  =\frac{2 \lsa}{L}\Big[1-\exp\Big(-\frac{L}{2\lsa}\Big)\Big]\,.
\end{align} 
This implies that
\begin{equation}
\label{eq:P1P0_E}
\frac{P_1}{P_0} = \frac{L}{2 g} \langle P_{\rm no\; overlap}\rangle = 
\frac{ \lsa}{g}\Big[1-\exp\Big(-\frac{L}{2\lsa}\Big)\Big] \,.
\end{equation}
This is Eq.~(\ref{eq:P1P0}) in the main text.

 \begin{figure}[b]
\mbox{}\hspace*{-8cm}
 \includegraphics[width=8cm]{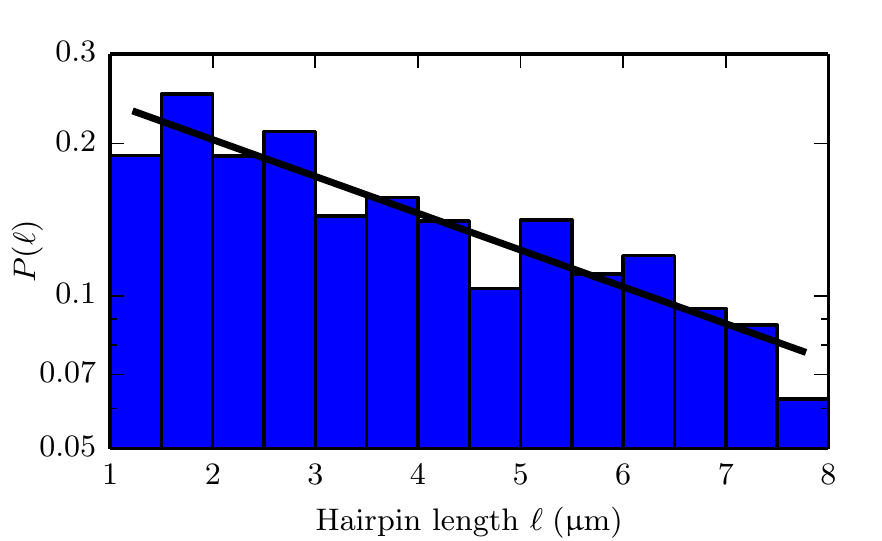}
 \caption{\label{fig:hairpinLengthDistribution} Semi-logarithmic plot of the probability distribution of the hairpin length for a polymer with persistence length $\ellp= 1150$ nm, in a channel of width $\dw = 600$ nm. The solid line indicates an exponential fit to Eq.~(\ref{eq:lsa}), leading to $\lsa=6.0\, \upmu$m. Together with the numerical values for $g$ and $\xi$ (see Table~\ref{tab:2}), this gives us  a dimensionless prefactor of $\alpha = 1.7$ after comparing against Eq.~(\ref{eq:P_noOverlap}). }
\end{figure}

\section{Estimation of $\lsa$}
\label{app:B}

We determined the self-avoidance length scale $\lsa$ from the configurations produced by the PERM simulations using  a polymer with a persistence length of $\ellp = 1150\,$nm confined to a channel of width $\dw = 600$ nm. To compute this data, we used $3 \times 10^5$ tours, corresponding to $2.7 \times 10^6$ configurations. For each tour that reached the maximum contour length, we output the configuration of the chain and its corresponding statistical weights for post-processing. The configuration of a polymer chain allows us to calculate the hairpin-contour lengths. To do so, we considered a moving window of 10 beads and computed the span and end-to-end distance for the subchain in a given window. If these two sizes were found to be equal, we concluded that there was no hairpin within the window; otherwise, the window was flagged as belonging to a hairpin. Typically, a given hairpin involves multiple contiguous windows, which were grouped into a single hairpin event by a clustering algorithm. Within each cluster, we then identified the bead with the minimum value of position $x_i$ along the channel axis and assigned the hairpin location to that bead. If the algorithm identified two or more hairpin strands, we used the shortest strand length as $\ell$. A naive application of this algorithm produces many small hairpins that would be undetectable by the optical resolution in the experiments. To provide a result that is experimentally relevant, we only included  values of $\ell$ if the chain within that cluster spans wider than 500 nm.

We extracted the self-avoidance length scale $\lsa$ from the distribution of $\ell$ using Eq.~(\ref{eq:P_noOverlap}). Figure \ref{fig:hairpinLengthDistribution} shows the probability distribution of hairpin lengths for a self-avoiding polymer in a channel. The value of $\lsa$ is obtained by fitting the distribution to Eq.~(\ref{eq:P_noOverlap}); as seen in Fig.~\ref{fig:hairpinLengthDistribution}. 
Since the hairpin length of an ideal polymer is uniformly distributed between $0$ and $L/2$, it follows that $P(\ell)$ equals $(2/L) P_{\rm no  \,overlap}(\ell)$. The parameter $\alpha$ must be independent of the channel size, so we used the result from Fig.~\ref{fig:hairpinLengthDistribution}, $\alpha=1.7$, to compute the values of $\lsa$ in Table \ref{tab:2}.

\section{Estimation of $\ellp$}
\label{app:C}
\begin{figure}
\mbox{}\hspace*{-8cm}
 \begin{overpic}[width=7cm]{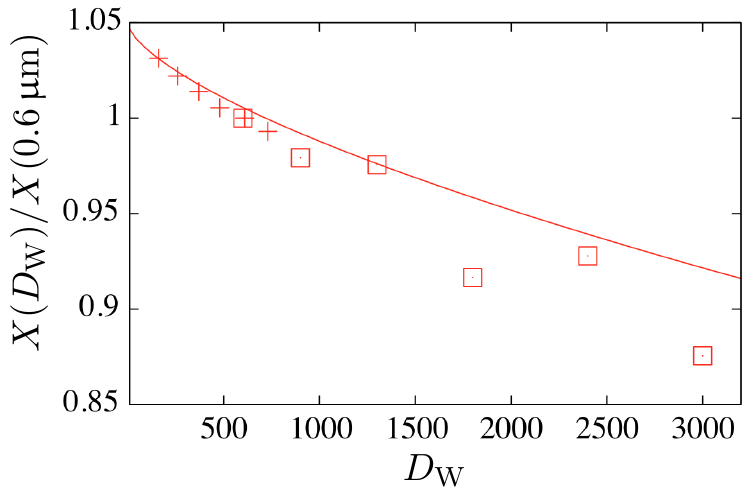}
\end{overpic}
\caption{\label{fig:small}
Median span taken over all DNA-RecA filaments from our experiments ($\Box$), normalized
by $X_{\rm median}(600\,{\rm nm})$. Experimental data ($+$) read off from Fig.~2A in Ref.~\citenum{frykholm2014} corresponding to measurements on a single DNA-RecA filament, normalized by
the span at $\dw=610\,{\rm nm}$.
 Also shown is Eq.~(\ref{eq:fry}) for $\ellp= 2\,\upmu$m and $L=13.78\,\upmu$m (solid line).}
\end{figure}

Fig.~\ref{fig:small} shows the same experimental data for the median of the spans of all DNA-RecA filaments as Fig.~\ref{fig:determine_lP} in the main text. In addition also the experimental data from Fig.~2A in Ref.~\citenum{frykholm2014} are shown, corresponding to measurements on a single DNA-RecA filament. Fig.~\ref{fig:small}  indicates that the data are consistent with each other.  Also shown is Odijk's theory, Eq.~(\ref{eq:Odijk_mean}),
\begin{equation}
\label{eq:fry}
\langle X \rangle = L\,\Big\{1-0.091\Big[\Big(\frac{\dhh}{\ellp}\Big)^{2/3} + \Big(\frac{\dw}{\ellp}\Big)^{2/3}\Big]\Big\}
\end{equation}
using $L=13.78\,\upmu$m and $\ellp=2\,\upmu$m. In Ref.~\citenum{frykholm2014} the same data were fitted with 
$L\approx 13.6\,\upmu$m, $\ellp \approx 1.39\,\upmu$m, and using a numerical prefactor $0.085$ that applies to Gaussian confinement rather than the value $0.091$ appropriate for rectangular channels. This yields a better fit at larger values of $\dw$, but one must bear in mind that Eq.~(\ref{eq:fry}) is only asymptotically valid at small $\dw/\ellp$. We can conclude that the fit of Eq.~(\ref{eq:fry}) to the experimental data is relatively insensitive to $\ellp$, in particular if $L$ must be fitted at the same time. A further source of uncertainty is that the variance between estimates of $\ellp$ obtained in Ref.~\citenum{frykholm2014} from different filaments is large, and that the average value $\ellp = 1.15\,\upmu$m obtained in Ref.~\citenum{frykholm2014} over many filaments does not include filaments with persistence lengths larger than $2\,\upmu$m. 
We also note that outliers in the data
were discarded, possibly leading to a bias to smaller persistence lengths
(the analysis is described in detail in Ref.~\citenum{frykholm2014}  and its supplementary information).
For these reasons we have chosen to determine $\ellp$ by comparison with simulation result in the range of (larger) $\dw$ directly relevant for our study. This results in $\ellp=2\,\mu$m, but also with a large uncertainty because the minimum in Fig.~\ref{fig:determine_lP} (b) is so shallow.

\end{document}


\title{Supplemental material for ``Emergence of hairpins in the conformations of a confined polymer"}
\author{E. Werner$^1$, A. Jain$^2$, A. Muralidhar$^2$, K. Frykholm$^3$, T. St Clere Smithe$^1$, J. Fritzsche$^4$, F. Westerlund$^3$, K. D. Dorfman$^2$, B. Mehlig$^1$}
\affiliation{\mbox{}$^1$Department of Physics, University of Gothenburg, Origov\"{a}gen 6B, 412 96 G\"{o}teborg, Sweden}
\affiliation{\mbox{}$^2$Department of Chemical Engineering
and Materials Science, University of Minnesota -- Twin Cities,
421 Washington Avenue SE, Minneapolis, Minnesota 55455, USA}
\affiliation{\mbox{}$^3$Department of Biology and Biological Engineering, Chalmers University of Technology, 412 96 G\"{o}teborg, Sweden}
\affiliation{\mbox{}$^4$Department of Physics, Chalmers University of Technology, Kemig{\aa}rden 1, 412 96 G\"{o}teborg, Sweden}
\maketitle

\noindent \section*{Kymographs - graphical representations of the experimental data} 
In this Supplementary Material we provide graphical representations of all the experimental videos, in the form of \lq kymographs\rq{}  (see Section II A of the main text).  The kymographs  were produced by stacking into a column the intensity profiles (below) produced for each sequence of video frames, so that each row of the column represents a single frame of the experimental recording.  

To produce the intensity profiles, the section of the frame containing the molecule was identified by locating the region with the maximal brightness. Next, the pixel intensity was averaged over the direction perpendicular to the channel, resulting in a row of pixel values which we interpret as the intensity profile along the channel.  The kymographs of all recorded molecules are listed below. Note that not all molecules have been recorded in all channels.
The molecule identifier includes the following information (from left to right):
\begin{enumerate}
\item Acquisition date.
\item Polymer type, either RecA-T4 or RecA-lambda,
where T4 and lambda refer to the type of DNA used for filament assembly.
\item The order in which the molecule traversed the different channels. The label \lq wide\rq{} indicates that the videos were recorded in order of decreasing channel sizes, \lq narrow\rq{} that the videos were recorded in order of increasing channel size.
\item A single number to uniquely identify each molecule.
\end{enumerate}
\clearpage
\newpage

\includepdf[angle=270, pagecommand={}, pages=-]{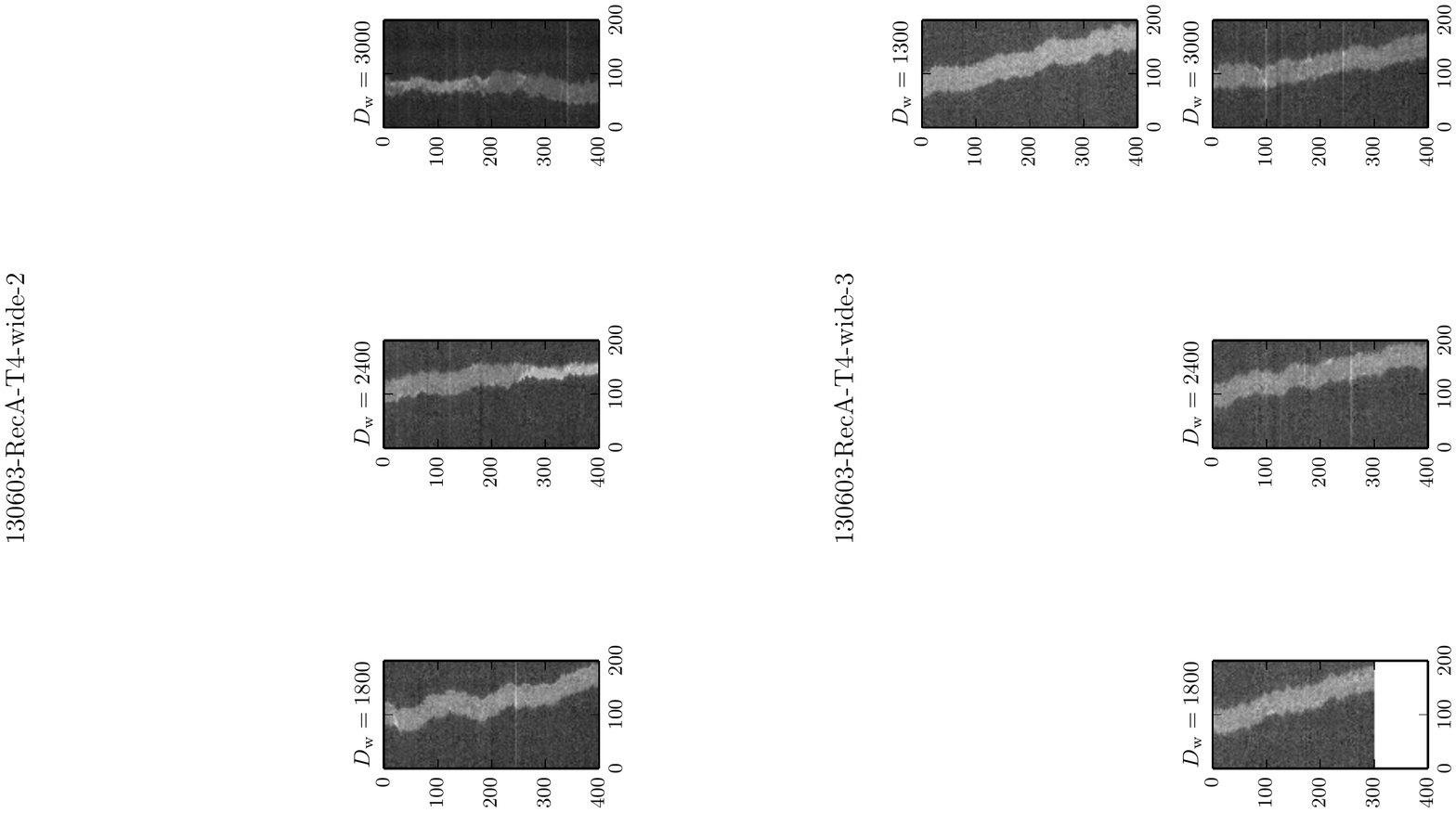} 

